\documentclass[preprint2]{aastex}
\usepackage{natbib}
\usepackage{wasysym}
\usepackage{color}
\usepackage{longtable}
\usepackage{booktabs}
 \usepackage{multirow}

\shorttitle{ENERGY LEVELS AND TRANSITION RATES OF N-LIKE IONS}
\shortauthors{Wang et al.}

\begin{document}


\title{Calculations with spectroscopic accuracy: energies and transition rates in the nitrogen isoelectronic sequence from Ar XII to Zn XXIV
}

\author{K. Wang\altaffilmark{1,2,3}, R. Si\altaffilmark{3,4},  W. Dang\altaffilmark{1}, P. J\"{o}nsson\altaffilmark{5}, X.L. Guo\altaffilmark{3,4}, S. Li\altaffilmark{2,3,4},   Z.B. Chen\altaffilmark{6}, \\H. Zhang\altaffilmark{2}, F. Y. Long\altaffilmark{2}, H.T. Liu\altaffilmark{2}, D.F. Li\altaffilmark{2},   R. Hutton\altaffilmark{3,4}, C.Y. Chen\altaffilmark{3,4}, and J. Yan\altaffilmark{2,7,8}}


\email{chychen@fudan.edu.cn}
\email{yan$_{-}$jun@iapcm.ac.cn}

\affil{$^1$Hebei Key Lab of Optic-electronic Information and Materials, The College of Physics Science and Technology, Hebei University, Baoding 071002, China}
\affil{$^2$Institute of Applied Physics and Computational Mathematics, Beijing 100088, China}
\affil{$^3$Applied Ion Beam Physics Laboratory, Fudan University, Key Laboratory of the Ministry of Education, China}
\affil{$^4$Shanghai EBIT Lab, Institute of Modern Physics, Department of Nuclear Science and Technology, Fudan University, Shanghai 200433, China}
\affil{$^5$Group for Materials Science and Applied Mathematics, Malm\"o University, S-20506, Malm\"o, Sweden}
\affil{$^6$College of Science,  National University of Defense Technology, Changsha 410073, China}
\affil{$^7$Center for Applied Physics and Technology, Peking University, Beijing 100871, China}
\affil{$^8$Collaborative Innovation Center of IFSA (CICIFSA), Shanghai Jiao Tong University, Shanghai 200240, China}

\begin{abstract}
Combined relativistic configuration interaction and many-body perturbation calculations are performed for the 359 fine-structure levels of the $2s^2 2p^3$, $2s 2p^4$, $2p^5$, $2s^2 2p^2 3l$, $2s 2p^3 3l$, $2p^4 3l$, and $2s^2 2p^2 4l$ configurations in N-like ions from \ion{Ar}{12} to \ion{Zn}{24}. 
A complete and consistent data set of energies, wavelengths, radiative rates, oscillator strengths, and line strengths for all possible electric dipole, magnetic dipole, electric quadrupole, and magnetic quadrupole transitions among the 359 levels are given for each ion. 
The present work significantly increases the amount of accurate data for ions in the nitrogen-like sequence, and the accuracy of the energy levels is high enough to serve identification and interpretation of observed spectra involving the $n=3,4$ levels, for which the experimental values are largely scarce. 
Meanwhile, the results should be of great help in modeling and diagnosing astrophysical and fusion plasmas.

\end{abstract}





\section{INTRODUCTION}\label{sect:in}
Spectra from L-shell ions, in a wide wavelength range from the X-ray to the ultraviolet, have been obtained from the solar atmosphere, stars, and other astronomical objects by many astrophysical missions, such as the \emph{Solar and Heliospheric Observatory}, \emph{Hinode}, \emph{Chandra}, and  \emph{Solar Dynamics Observatory}~\citep{Brinkman.2000.V530.p111,Landi.2002.V139.p281,Raassen.2002.V389.p228,Curdt.2004.V427.p1045,Ishibashi.2006.V644.p117,Brown.2008.V176.p511,DelZanna.2008.V481.p69,DelZanna.2012.V537.p38,Warren.2008.V685.p1277,DelZanna.2011.V528.p139,DelZanna.2013.V555.p59,Beiersdorfer.2014.V788.p25,Traebert.2014.V215.p6,Traebert.2014.V211.p14}. The analysis of the observed spectra provides information on the structure, chemical abundances, evolution, and physical conditions of the astrophysical objects. Such an analysis requires a wide range of atomic parameters, such as energy levels and radiative transition properties~\citep{Kallman.2007.V79.p79,Massacrier.2012.V538.p52,DelZanna.2013.V555.p59,Beiersdorfer.2014.V788.p25,Nave.2015.V.p}. 
In view of this, we have already reported the systematic and highly accurate calculations for the beryllium and carbon isoelectronic sequences~\citep{Wang.2014.V215.p26,Wang.2015.V218.p16}. 
This work presents our effort on the nitrogen isoelectronic sequence from \ion{Ar}{12} to \ion{Zn}{24}. Numerous lines of the above N-like ions have been observed in astrophysical plasmas, as well as in laboratory plasmas~\citep{Feldman.1980.V238.p365,Feldman.1997.V113.p195,McKenzie.1980.V241.p409,Doschek.1981.V245.p315,Eidelsberg.1981.V43.p455,Phillips.1982.V256.p774,Phillips.1983.V265.p1120,Doschek.1984.V56.p67,Lawson.1984.V58.p475,Acton.1985.V291.p865,Seely.1986.V304.p838,Doyle.1987.V173.p408,Fawcett.1987.V225.p1013,Lippmann.1987.V316.p819,Brosius.1998.V119.p255,Feldman.2000.V544.p508,Mewe.2001.V368.p888a,Curdt.2001.V375.p591,Behar.2001.V548.p966,Brown.2002.V140.p589,Kaastra.2002.V386.p427,Ko.2002.V578.p979,Lepson.2003.V590.p604,Mohan.2003.V582.p1162,Ness.2003.V598.p1277,Curdt.2004.V427.p1045,Feldman.2004.V607.p1039,Landi.2005.V160.p286,Landi.2006.V166.p421,Parenti.2005.V443.p679,Chen.2007.V168.p319,DelZanna.2008.V481.p69,DelZanna.2012.V537.p38,Gu.2007.V657.p1172,Shestov.2008.V34.p33,Beiersdorfer.2014.V788.p25,Traebert.2014.V215.p6,Traebert.2014.V211.p14}.

Many theoretical efforts have been devoted to studying energy levels and transition characteristics in N-like ions. Most of the systematic calculations, such as  ~\citet{Godefroid.1984.V17.p681,Becker.1989.V221.p375,Merkelis.1997.V56.p41,Merkelis.1999.V59.p122,Gu.2005.V89.p267} and~\citet{Rynkun.2014.V100.p315},  were limited to a few transitions among the 15 fine-structure levels of the $(1s^2) 2s^2 2p^3$, $2s 2p^4$, and $2p^5$ configurations (the $n=2$ complex).

To our knowledge, there were no systematic calculations beyond the $n = 2$ states along the isoelectronic sequence, except for one calculation preformed by~\citet{Tachiev.2002.V385.p716}. Using the multiconfiguration Hartree-Fock method with relativistic corrections in the Breit-Pauli approximation (MCHF-BP), they computed energies and  transition data for the levels up to $2s^2 2p^2 3d$ in N-like ions with $Z = 7-17$. However, a few calculations have been carried out for selected individual ions. \citet{Bhatia.1989.V43.p99,Eissner.2005.V89.p139} and \citet{Landi.2005.V90.p177} reported both the $n=2$ and $n=3$ results of \ion{Ar}{12}, \ion{Ca}{14}, \ion{Ti}{16}, \ion{Fe}{20}, \ion{Zn}{24}, and \ion{Kr}{30} using the SUPERSTRUCTURE (SS) code~\citep{Eissner.1974.V8.p270}.  
The energy levels and radiative decay rates for the transitions involving the $n \geq 3$ levels in \ion{Fe}{20} were calculated using various methods. {The calculations include the Breit–Pauli R-matrix (BPRM) calculations and the configuration interaction calculations using the SS code by \citet{Nahar.2004.V413.p779}, the multiconfiguration Dirac-Hartree-Fock (MCDHF) calculations by \citet{Jonauskas.2005.V433.p745}, the results of \citet{Witthoeft.2007.V466.p763}  using the AUTOSTRUCTURE (AS) code~\citep{Badnell.1986.V19.p3827}.
~\citet{Dong.2012.V64.p131} employed the MCDHF method in the GRASP package~\citep{Dyall.1989.V55.p425} to calculate level energies and radiative rates among the transitions for the 272 levels of the $n=2,3$ levels in \ion{Ca}{14}.  
Energy levels and radiative data for the transitions up to the $n=10$ levels in \ion{Sc}{15} were provided by~\citet{Massacrier.2012.V538.p52} using the FAC code~\citep{Gu.2003.V582.p1241,Gu.2008.V86.p675}. 
A combined relativistic configuration interaction (RCI) and many-body perturbation theory (MBPT) approach was used by \citet{Gu.2005.V156.p105} to obtain the level energies for the $2l^5$ and $2l^4 3l^\prime$ configurations in \ion{Fe}{20} and \ion{Ni}{22} with high accuracy. 

Among the above calculations for N-like ions, the results for the $n=2$ states reported by \citet{Rynkun.2014.V100.p315} and \citet{Gu.2005.V89.p267}, and the data for the $n=2,3$ levels obtained by \citet{Gu.2005.V156.p105}  in \ion{Fe}{20} and \ion{Ni}{22} are so far the most accurate. In contrast with the accurate values  \citep{Rynkun.2014.V100.p315,Gu.2005.V89.p267,Gu.2005.V156.p105}, all the other mentioned calculations involving the $n \geq 3$ complexes for highly charged N-like ions from \ion{Ar}{12} to \ion{Zn}{24} are quite inaccurate due to limited configuration interaction effects included in their works. For instance, the energies of the SS calculations \citep{Bhatia.1989.V43.p99,Eissner.2005.V89.p139,Landi.2005.V90.p177} for five ions  from \ion{Ar}{12} to \ion{Zn}{24} deviate from the corresponding observations by up to 5\%, which may be outdated for line identification and plasma diagnostics. 
In terms of theoretical works, \ion{Fe}{20} is currently the most studied ion in the  nitrogen isoelectronic sequence so far. The deviations from the observed energies are up to 3.4\% for the BPRM calculations~\citep{Nahar.2004.V413.p779},  2.2\% for the MCDHF values~\citep{Jonauskas.2005.V433.p745}, and 4.3\% for the AS results~\citep{Witthoeft.2007.V466.p763}, which are far from the spectroscopic accuracy. Therefore, systematic  calculations of high quality involving states beyond the $n=2$ configurations are greatly desired, because of their importance in modeling and diagnosing of  astrophysical plasmas \citep{Phillips.1982.V256.p774,Acton.1985.V291.p865,DelZanna.2008.V481.p69,Beiersdorfer.2014.V788.p25} and laboratory plasmas \citep{Fawcett.1975.V170.p185}. Databases such as CHIANTI \citep{Dere.1997.V125.p149,Landi.2013.V763.p86} also demand complete and consistent data sets of high accuracy, with the view of offering the astrophysical community tools and data to carry out accurate plasma diagnostics.

Recently, \citet{Radziute.2015.V582.p61} reported calculated energies and radiative transition properties for the 272 states of the $2s^2 2p^3$, $2s 2p^4$, $2p^5$, $2s^2 2p^2 3l$, $2s 2p^3 3l$, and $2p^4 3l$ ($l = 0, 1, 2$) configurations in N-like ions \ion{Cr}{18}, \ion{Fe}{20}, \ion{Ni}{22} and \ion{Zn}{24} by using the MCDHF and RCI method implemented in the GRASP2K code~\citep{Jonsson.2013.V184.p2197,Jonsson.2007.V177.p597}. Comparing with the calculations of \citet{Rynkun.2014.V100.p315} who used the same method while only reported the results for the $n=2$ complex, \citet{Radziute.2015.V582.p61} adopted much larger configuration state function expansions and considered the electron correlation effects elaborately for both the $n=2$ and $n=3$ levels. Therefore, high accuracy was achieved in their calculations, which was in general at the same level with the accuracy of the calculations performed by \citet{Rynkun.2014.V100.p315} and \citet{Gu.2005.V89.p267,Gu.2005.V156.p105}, and the data can be used to identify observed spectral lines.

In the present work, we report energy levels and transition properties for all possible electric dipole (E1), magnetic dipole (M1), electric quadrupole (E2), and magnetic quadrupole (M2) transitions among the 359 levels of the $2s^2 2p^3$, $2s 2p^4$, $2p^5$, $2s^2 2p^2 3l$, $2s 2p^3 3l$, $2p^4 3l$, and $2s^2 2p^2 4l$ configurations in the N-like ions with $18 \leq Z \leq 30$, in an effort to offer complete and consistent data sets of high accuracy. A combined RCI and MBPT approach implemented in the FAC code  \citep{Gu.2003.V582.p1241,Gu.2005.V89.p267,Gu.2005.V156.p105,Gu.2006.V641.p1227} is used, in which both dynamic and nondynamic electron correlation effects could be well accounted for. 
For the purpose of assessing the present MBPT results, extensive MCDHF and RCI calculations (hereafter referred to as MCDHF/RCI) for \ion{Fe}{20} have been carried out  using the latest version of GRASP2K code~\citep{Jonsson.2013.V184.p2197}. 
Comparisons are made between the present MCDHF/RCI and MBPT results, as well as with available observed data and theoretical values. The MBPT calculated energies agree well with the observed values from the Atomic Spectra Database (ASD) of the National Institute of Standards and Technology (NIST)~\citep{Kramida2014}, i.e., a difference is within 0.2\% for all levels. The present calculations are generally more accurate than existing systematic calculations, and stand for a significant extension of the MBPT work reported by \citet{Gu.2005.V156.p105} and the MCDHF/RCI results performed by \citet{Radziute.2015.V582.p61}  to include data for the other nine ions in the range of \ion{Ar}{12} to \ion{Zn}{24}. We hope that the present data should be of great help in analyzing older experiments and planning new ones. Meanwhile, complete data sets will be useful in the identification of observed spectra, as well as in modeling and diagnosing of astrophysical and fusion plasmas.

\section{Calculations and Results}\label{sect:cp}
A combined RCI and MBPT approach \citep{Lindgren.1974.V7.p2441,Safronova.1996.V53.p4036,Vilkas.1999.V60.p2808} was implemented within the FAC code by \citet{Gu.2005.V89.p267,Gu.2005.V156.p105}. In the present work, we employ the improved implementation by \citet{Gu.2006.V641.p1227}, in which the Hamiltonian is taken to be the no-pair Dirac--Coulomb--Breit Hamiltonian $H_{\rm DCB}$. The key feature of the RCI and MBPT approach is to partition the Hilbert space of the system into two subspaces, i.e., a model space \emph{M} and an orthogonal space \emph{N}. The true eigenvalues of $H_{\rm DCB}$ can be obtained through solving the eigenvalue problem of a non-Hermitian effective Hamiltonian in the model space \emph{M}. The first-order perturbation expansion of the effective Hamiltonian within the Rayleigh--Schr{\"o}dinger scheme consists of two parts: one is the exact $H_{\rm DCB}$ matrix in the model space \emph{M}, and the other includes perturbations from the configurations in the \emph{N} space up to the second order for the level energies of interest. In the present calculations, the model space \emph{M} contains all of the configurations $2l^4 n l^\prime$ ($2\leq n \leq 3$ and $l^\prime \leq n-1$) and $2s^2 2p^2 4l^\prime$ ($l^\prime=0-3$). The \emph{N} space contains all  configurations formed by single and double virtual excitations of the \emph{M} space. For single excitations, configurations with $n\leq 200$ and $l\leq \min{(n-1,25)}$ are included. For double excitations, configurations with the inner electron promotion up to $n = 65$ and promotion of the outer electron up to $n^\prime = 200$ are considered.

We start the energy structure calculations for N-like ions using an optimized local central potential, which is derived from a Dirac--Fock--Slater self-consistent field calculation with the $ (2s,2p)^5 $ configurations. We then perform the MBPT calculations to obtain level energies and radiative transition properties, such as transition wavelengths, line strengths, oscillator strengths and radiative rates of all E1, M1, E2, and M2 transitions  among the states in the \emph{M} space using the length form. In addition to the Hamiltonian $H_{\rm DCB}$, several high-order corrections, such as the finite nuclear size, nuclear recoil, vacuum polarization (VP), and electron self-energy (SE), are also taken into account in the calculations.

Table~\ref{tab.lev.sub} lists our energies(in eV) of the 359 levels for each ion, 
as well as the observed results from the NIST ASD~\citep{Kramida2014}.  Also listed in Table~\ref{tab.lev.sub} is the symmetry $J^{\pi}$ of the state, i.e.,  the parity $\pi$ and total angular momentum \emph{J}.  For each state in each ion, an index number \emph{j}/\emph{i} is assigned, which can be used in Table~\ref{tab.tr.sub} to represent the upper (\emph{j}) or lower (\emph{i}) levels of the radiative transition. The wavelength ($\lambda_{ji}$ in $\rm {\AA}$), line strength ($S_{ji}$ in atomic units, 1 AU = $\rm 6.460\times 10^{-36} cm^2 esu^2$), oscillator strength ($f_{ji}$ dimensionless) and radiative rate ($A_{ji}$ in s$^{-1}$) for all possible E1, M1, E2, and M2 transitions among the 359 levels, are listed in Table~\ref{tab.tr.sub}.

In a large-scale calculation, it is difficult to label unambiguously all of the states. Using the $LSJ$- or $jj$ coupling schemes, the identifications and the corresponding configurations can be determined clearly for a majority of the states listed in Table~\ref{tab.lev.sub}. However, in the symmetry of $\rm {3/2}^{o}$ in $2s^2 2p^2 np$ configurations and $\rm {3/2}^{e}$ in $2s^2 2p^2 nd$ configurations, the identifications of some states are not unique.  The dominant components of the configuration basis for each state in both the $LSJ$- and $jj$ coupling schemes are also given  for each state in Table~\ref{tab.lev.sub} to help to remove these ambiguities. The $jj$ notation is given directly by the MBPT calculations,  and the $LSJ$- notation is obtained using the $jj \rightarrow LSJ$- transformation presented by~\citet{Gaigalas.2004.V157.p239}.

For assessing the accuracy of the MBPT results,  extensive MCDHF and subsequent RCI calculations are performed for \ion{Fe}{20} using the GRASP2K code~\citep{Jonsson.2013.V184.p2197}. In the MCDHF method, the Hamiltonian is taken to be Dirac--Coulomb Hamiltonian $H_{\rm DC}$.
The atomic state functions (ASFs) are approximated by expansions over $jj$-coupled configuration state functions (CSFs).
Based on equal level weights of several states, the so called extended optimal level (EOL) scheme
~\citep{Dyall.1989.V55.p425}, both the radial parts of the Dirac orbitals and the expansion coefficients were optimized to self-consistency in the relativistic self-consistent field procedure.
In the subsequent RCI calculation~\citep{McKenzie.1980.V21.p233}, the Breit interaction is computed in the low-frequency limit by multiplying the frequency with a scale factor of $10^{-6}$, and the other small corrections such as finite nuclear size, nuclear recoil, VP and SE, are also included.

In the present MCDHF/RCI calculations, the configurations in the \emph{M} space of the above MBPT calculations are split up into the even and odd groups,  and are chosen as the reference configurations. Then the multi-configuration expansions are obtained through single and double excitations of the orbitals in the reference configurations with orbitals in an active set with principal quantum numbers $n = 3,...,7$ and angular symmetries $s$, $p$, $d$, $f$, $g$, and $h$. To monitor the convergence of the calculated energies and transition parameters, the active sets were increased in a systematic way by adding layers of orbitals. For the $n = 7$ expansion this resulted in 1182541 CSFs with odd parity and 1150043 CSFs with even parity. The self-consistent field calculations for each layer of orbitals are followed by the RCI calculations. A more detailed description of the calculation procedure could be found in our very recent work for Mg-like  \ion{Cu}{18} and \ion{Kr}{25}~\citep{Si.2015.V48.p175004,Si.2015.V163.p7}.

\section{Comparisons}\label{sect:com}
Employing the MBPT approach, we have carried out systematic and highly accurate calculations for energy levels and transition rates of the beryllium and carbon isoelectronic sequences~\citep{Wang.2014.V215.p26,Wang.2015.V218.p16}. We now focus our attention on the nitrogen isoelectronic sequence, and our work greatly increases accurate results for nitrogen-like ions in quantity. The accuracy of our MBPT results will be accessed, in the following,  by comparing with available observed values, as well as with other elaborate/systematic calculations.

\subsection{\emph{Energy Levels}}\label{sect:en}
Our MBPT values agree well with the NIST observations for all the 182 $n=2$ states of the thirteen ions considered here, i.e., the differences are within 0.1\% for 157 levels, and are between 0.1\% and 0.2\% for the remaining 25 levels.  
The previous theoretical results, including the MBPT calculations reported by   \citet[hereafter referred to as MBPT2]{Gu.2005.V89.p267} and the MCDHF and RCI calculations performed by \citet[MCDHF/RCI2]{Rynkun.2014.V100.p315}, are so far the most accurate in the sequence, when compared with the observed values.  
 The relative differences between calculations and observations are within 0.1\% for 150 levels in MBPT2, and 138 levels in MCDHF/RCI2. The deviations are larger than 0.2\% for 7 levels (up to 0.3\% for $2s^2 2p^3\ {}^{2}\!P_{3/2}$ in \ion{Ar}{12}) in MBPT2, and  14 levels (up to 0.34\% for $2s^2 2p^3\ {}^{2}\!D_{3/2}$ in \ion{Ca}{14}) in MCDHF/RCI2. Comparing with the MBPT2 and MCDHF/RCI2 calculations, a general better agreement of the NIST values and our results is attained due to more effectively electron correlation effects included within our work by considering more configurations in both the \emph{N} and \emph{M} spaces.
 
Recently, using the MCDHF and RCI  method \citet{Radziute.2015.V582.p61} reported the energies and transition rates for the 272 states of the $(2s,2p)^5$ and $ (2s,2p)^4 3l$ ($l = 0, 1, 2$) configurations in N-like ions \ion{Cr}{18}, \ion{Fe}{20}, \ion{Ni}{22}, and \ion{Zn}{24} (MCDHF/RCI3). Since effective  configuration interaction effects have been taken into account, their data of high accuracy can be used to identify the observed spectra. Comparisons of the NIST and CHIANTI experimental results,  the present calculations, and the MCDHF/RCI3 values for the $n=2$ states in \ion{Fe}{20} show good agreements among them, i.e., the mean relative deviations from NIST \& CHIANTI experimental values are 0.021\% \& 0.021\% for this work, 0.041\% \& 0.046\% for the MCDHF/RCI3 results; plus the mean (with standard deviation) of the relative differences between the two sets of calculated values is 0.015\%$\pm$0.015\% for all 272 states of the $n=2,3$ configurations, which, again, is highly satisfactory. 

The experimental values are sparse for the $n = 3,4$ levels of N-like ions. To confirm the accuracy of the present $n \geq 3$ results and show the importance of the electron correlation effects, we have performed another independent calculation for \ion{Fe}{20} using the MCDHF and RCI method. In Table~\ref{tab.lev.FeXX}, energies for the 344 levels of the $n=3,4$ complexes in \ion{Fe}{20} from the MBPT and MCDHF/RCI calculations in this work are compared with calculated values from a previous calculation using the AS code by~\citet{Witthoeft.2007.V466.p763}. 
Collected in Table~\ref{tab.lev.FeXX} are also experimental values from the NIST and CHIANTI databases. The deviations of the MCDHF/RCI  and MBPT energies are plotted in Figure~\ref{fig.lev.FeXX}. It can be seen that the agreement between the two calculations is better than 0.08\% for all the 344 levels. The mean (with standard deviation) of the relative differences between the two sets of energy values is 0.023\%$\pm$0.012\%, which is highly satisfactory. 

As shown in Table~\ref{tab.lev.FeXX}, experimental observations are largely missing, i.e., the NIST \& CHIANTI databases list the energies for 27 \& 37 out of the  344 $n=3,4$ levels in \ion{Fe}{20}. 
Furthermore,  the identification of these levels becomes problematic in some cases. 
For example, the $2s^2 2p^2 (^3P) 3d \   {}^{4}P_{5/2 }$ (967.32 eV) level~\citep{Sugar.1985.V14.p1} from the NIST ASD is observed at a considerably higher energy ($>$2.1 eV) than from the CHIANTI database and the present MBPT and MCDHF/RCI calculations.  
The $2s^2 2p^2 (^3P) 3d \   {}^{2}D_{3/2 }$ (974.39 eV), $2s^2 2p^2 (^1D) 3d \   {}^{2}F_{5/2 }$ (989.77 eV), $2s^2 2p^2 (^3P) 4d \   {}^{2}D_{3/2 }$ (1245.17 eV), and $2s^2 2p^2 (^1S) 4d \   {}^{2}D_{5/2 }$ (1275.67 eV) levels in the NIST ASD are also higher than the present MBPT and MCDHF/RCI calculations by over 2-3 eV. 
The misidentification for these five levels from the NIST ASD cannot be ruled out, as suggested by~\citet{Radziute.2015.V582.p61}.
The relative differences between the NIST observations and the MBPT calculations (or the MCDHF/RCI calculations) are within 0.2\% for the remaining 22 $n=3,4$ levels in \ion{Fe}{20}. 
The agreement with the CHIANTI observed values is better.   
Deviations of the CHIANTI values relative to the MBPT results (or the MCDHF/RCI results) are less than 0.2\% for 33 levels, except for the $2s 2p^3 (^3S) 3p  \   {}^{4}P_{5/2 }$ level. The  CHIANTI value is 1056.17 eV for the ${}^{4}P_{5/2 }$ level, which is very close to our MBPT and MCDHF/RCI results for the $2s 2p^3 (^1D) 3p  \   {}^{2}F_{5/2 }$ level, 1056.83 and 1056.99 eV. This indicates that the CHIANTI results should perhaps be designated as the $2s 2p^3 (^1D) 3p  \   {}^{2}F_{5/2 }$ level. The mean differences with the CHIANTI values are both 0.05\% for the MBPT and MCDHF/RCI calculations. The other theoretical energies shown in Table~\ref{tab.lev.FeXX}, i.e., the  AS results, differ from the CHIANTI values significantly. The deviations of the AS results from the CHIANTI values are up to 0.42\%  for the $2s^2 2p^2(^3P)3s \   {}^{4}P_{5/2 }$ levels. The mean deviation relative to the CHIANTI values is 0.24\%, being larger by over one order of magnitude than the present MBPT (MCDHF/RCI)  results. 


The NIST ASD only includes 108 levels of the total 4472 $n \geq 3$ levels in the 13 N-like ions considered here. The relative deviations between the MBPT and NIST values are less than 0.2\% for 85 levels. The remaining 23 states for which the differences are larger than 0.2\% are listed in Table~\ref{tab.lev.nistmbptlardif}.  
First, we look at the states for which the differences between the  MBPT and NIST values exceed 0.5\%.  The NIST energy 404.004 eV for the $2s^2 2p^2 (^1D) 3d  \   {}^{2}G_{7/2}$ level is very close to our MBPT result for the $2s^2 2p^2 (^1D) 3d  \   {}^{2}F_{7/2}$ level, 403.552 eV, which suggests that it might be mislabeled and should be designated as the ${}^{2}F_{7/2}$ level. Figure~\ref{fig.lev.nistlargedifferences}(a) display the deviations for two representative states ($2s^2 2p^2 (^1D)3d\/~{}^2\!D_{5/2}$ and $2s^2 2p^2 (^1S)3d\/~{}^2\!D_{5/2}$), as well as the differences for the same states but in other N-like ions so as to aid in comparisons. The deviations are 0.39\%, 0.45\%, and 0.65\% for the $2s^2 2p^2 (^1D)3d\/ ~{}^2\!D_{5/2}$ level in \ion{K}{13}, \ion{Ca}{14}, and \ion{V}{17}, respectively. Whereas the MBPT
results agree well with the observations for the same level in \ion{Sc}{15} and \ion{Fe}{20}.  In fact, the NIST energy for  the $2s^2 2p^2 (^1D)3d\/ ~{}^2\!D_{5/2}$ level in \ion{K}{13} is 463.2 eV, and the observed wavelengths 27.53  ($2s^22p^3\ ^2D_{5/2} - 2s^22p^2(^1D)3d\ ^2D_{5/2}$) and  28.012  ($2s^22p^3\ ^2P_{3/2} - 2s^22p^2(^1D)3d\ ^2D_{5/2}$) \AA ~are used to deduce the energy for $2s^22p^2(^1D)3d\ ^2D_{5/2}$ with the aid of the known energies of $2s^22p^3\ ^2D_{5/2}$ and $^2P_{3/2}$~\citep{Fawcett.1975.V170.p185}. The MBPT wavelengths, respectively,  are 27.656 and 28.142 \AA~for the above two transitions, which deviate from the observed values (27.53 and  28.012 $\AA$) up to 0.46\% and 0.47\%, but agree well (0.08\% and 0.15\%) with the more recent experimental values 27.63 and 28.10 \AA~in the revised identification recommended by the same group~\citep{Bromage.1977.V179.p683}.  Using the above revised observations, the experimental data for $2s^22p^3\ ^2D_{5/2}$ should be corrected as 461.8  and 462.1 eV, respectively,  with the aid of the known energies of $2s^22p^3\ ^2D_{5/2}$ and $^2P_{3/2}$, which agree with our MBPT value of 461.403 eV to within 0.15\%. The NIST values of the $2s^22p^3\ ^2D_{5/2}$ level in \ion{Ca}{14}, and \ion{V}{17} are also determined using the observations given by~\citet{Fawcett.1975.V170.p185}. We consider that the observation uncertainty should be relatively large according to similar arguments for \ion{K}{13} used above. Furthermore, our MBPT energies of the $2s^22p^2(^1D)3d\ ^2D_{5/2}$ level in eight ions (from \ion{K}{12} to \ion{Fe}{20})  agree well (within 0.1\%) with the calculated values recommended by \citet{Bromage.1977.V179.p683}. Thus, we can state with confidence that the observation uncertainty for the $2s^22p^2(^1D)3d\ ^2D_{5/2}$ level in \ion{K}{13}, \ion{Ca}{14}, and \ion{V}{17} in the NIST ASD exceeds the "uncertainty" of the MBPT calculations. As also shown in Figure~\ref{fig.lev.nistlargedifferences}(a), the NIST and MBPT values agree within 0.01\% for the same level in \ion{Fe}{20}, but the NIST compiled energy of 605.0 eV for the $2s^2 2p^2 (^1S)3d\/~{}^2\!D_{5/2}$ state in \ion{Sc}{15} is  0.58\% higher than the MBPT values. 
This is quite strange because the unified treatment is adopted for each ion in the present MBPT calculations, which implies that the calculated results of \ion{Sc}{15} may only be slightly less accurate than those of \ion{Fe}{20} because of more important effects of the electron correlation in the relatively \emph{lower-Z} system in principle~\citep{ Wang.2015.V218.p16}. In addition, 
 ~\citet{Sugar.1985.V14.p1} said that the $2s^2 2p^2 (^1S)3d\/~{}^2\!D_{5/2}$ state in \ion{Sc}{15} was from the interpretation by \citet{Fawcett.1975.V170.p185} and \citet{Bromage.1977.V179.p683}. We look over the above two papers, but do not find the measurement regarding the $2s^2 2p^2 (^1S)3d\/~{}^2\!D_{5/2}$ state. Therefore, the NIST value for the $2s^2 2p^2 (^1S)3d\/~{}^2\!D_{5/2}$ level in \ion{Sc}{15} may not be determined by the measurement, and we argue that the uncertainty might be large for this level.

In Table~\ref{tab.lev.nistmbptlardif}, there are 17 levels for which the differences between the  MBPT and  NIST values are in the range of 0.2\%--0.5\%. In Figure~\ref{fig.lev.nistlargedifferences}(b) we show the comparison of some representative cases. In order to discuss the Z-dependent behavior, Figure ~\ref{fig.lev.nistlargedifferences}(b) also displays the energy deviations ($ <\!0.2$\%) for levels in other ions along the sequence. 
The NIST and MBPT values agree within 0.2\% for most levels, and large  differences happen at the \emph{higher-Z} end ($2s^2 2p^2(^1D)3d\/~{}^2\!F_{5/2}$ and $2s^2 2p^2 (^3P)3d\/~{}^2\!D_{3/2}$ and $~{}^4\!P_{5/2}$ ) and/or in the middle of the isoelectric sequence ($2s^2 2p^2 (^3P)3d\/~{}^2\!D_{3/2}$ and $~{}^2\!P_{3/2}$). Theoretically, this is not reasonable because the electron correlation effects are usually more significant for  \emph{lower-Z} ions. Furthermore, the present MBPT and MCDHF/RCI results agree well within 0.05\% for the $2s^2 2p^2 (^1D)3d\/~{}^2\!F_{5/2}$ and $2s^2 2p^2 (^3P)3d\/~{}^2\!D_{3/2}$ and $~{}^4\!P_{5/2}$ levels in \ion{Fe}{20}.
Therefore, we consider that the large deviations for the 23 $n=3, 4$ levels listed in Table~\ref{tab.lev.nistmbptlardif} should be mainly caused by relatively large observation uncertainty.

\subsection{\emph{Radiative rates}} 
In Figure~\ref{fig.tr.e1.n2.mcdf}, we compare the present MBPT line strengths (\emph{S}~values) with the MCDHF/RCI2 results \citep{Rynkun.2014.V100.p315} for  all of 637 E1 transitions among the $n=2$ levels in  N-like ions with $18 \leq Z \leq 30$. 
The transitions for which the \emph{S}~values $<\! 10^{-6}$ are not displayed in Figure~\ref{fig.tr.e1.n2.mcdf}, because there are only five transitions in this range.  
For most of the transitions (except for 6 relatively weak transitions with the \emph{S}~values $<\! 10^{-5}$), the agreement of the two sets of results is within 5\%, which is highly satisfactory. The differences of two data sets result from different electron correlation effects considered in the calculations. Weak transitions are generally very sensitive to the electron correlation effects considered in the calculations, and sometimes are even sensitive to the high-order relativistic effects which might be necessary to describe the coupling conditions very accurately.

The NIST ASD lists the \emph{S}~values for 337 out of the total 637 transitions among the $n=2$ complex. Figure~\ref{fig.tr.e1.n2.nist}  compares the present MBPT \emph{S}~values with those included in the NIST ASD. The NIST values differ from the MBPT results by over 10\% for 74\% out of the 337 E1 transitions. Even for many very strong transitions with the  \emph{S}~values $ \geq 10^{-2}$, the differences are also larger than 10\%. 
Table~\ref{tab.tr.e1.n2.nist.largediff} lists all the E1 transitions for which the deviations between the NIST and MBPT \emph{S}~values exceed 10\%.
We notice that these transitions belong to the ions with $21-28$, and the source of the NIST data is from \citet{Cheng.1979.V24.p111},  which may be out of date. In their calculations, the MCDHF technique was used to calculate energy levels and wave functions for the $n=2$ levels. Only the configurations within the $n = 2$ complex were included to account for electron correlations and intermediate coupling. As shown in Figure~\ref{fig.tr.e1.n2.nist} and  Table~\ref{tab.tr.e1.n2.nist.largediff}, there is a good agreement (within 2\%) between the present MBPT \emph{S}~values and those from \citet{Rynkun.2014.V100.p315},  The present MBPT calculations, as well as the MCDHF/RCI2 results, should be more accurate and reliable than those values recommended by the NIST ASD.

Of the large number of the E1 transitions involving the higher excited levels of the $n=3,4$ complexes in the 13 N-like ions, the NIST ASD only lists the \emph{S}~values for 26 one-photon-one-electron transitions in \ion{Fe}{20}. These NIST values are included in Table~\ref{tab.tr.FeXX.n3.nist}, as well as the present MBPT and MCDHF/RCI results. Even for the one electron transitions, the differences of the NIST values relative to the MBPT results (or the MCDHF/RCI results) are over 20\% (up to a factor of 7 for $2s^2 2p^2(^1D) 3d\ ^2D_{5/2}\rightarrow 2s^2 2p^3 \ ^4S_{3/2}$ for which $\Delta S = 1$ and $\Delta L = 2$}) in six cases. As seen from Table~\ref{tab.tr.FeXX.n3.nist}, the agreement between the present MBPT and MCDHF/RCI \emph{S}~values is satisfactory, being within 4\% for all 26 E1 transitions. To further assess the accuracy of the present transitions properties,  in Figure~\ref{fig.tr.mbpt.mcdf.FeXX} the  MCDHF/RCI \emph{S}~values are plotted against the MBPT results for  2307 strong transitions (\emph{S}~values $ \geq 10^{-2}$) in \ion{Fe}{20}.  
For 83\% of these strong transitions, our MBPT and MCDHF/RCI \emph{S}~values agree to within 5\%, while they differ from each other by over 20\% (but no more than a factor of 3.5) for  67 transitions. Many of these transitions are intercombination E1 transitions, for which cancellation effects often decrease the accuracy considerably. Nevertheless, except for those 67 transitions, the average difference (with standard deviation) between our MBPT and MCDHF/RCI \emph{S}~values for the remaining 2240 strong transitions is only $2\% \pm 4\%$.

The NIST ASD lists the \emph{S}~values for the total 129 M1 and E2 transitions in the 13 ions considered here, which are all within the $n=2$ states. The NIST values for these 129 transitions are compared with the MBPT results in Table~\ref{tab.tr.m1.e2.n2.nist.largediff}. The MCDHF/RCI2 results are also collected in Table~\ref{tab.tr.m1.e2.n2.nist.largediff}  for comparison. 
The deviations of the NIST \emph{S}~values relative to the present MBPT results
are less than 10\% for 109 transitions, are between 10\% and 30\% for 20 transitions. The largest discrepancy is about 27\% for the $2s 2p^4\ {}^{2}\!D_{3/2} \rightarrow \ {}^{4}\!P_{1/2}$ transition in \ion{Ti}{16}. The MBPT and MCDHF/RCI2 values agree well (1\%) for this transition.
The MCDHF/RCI2 calculations reported the \emph{S}~values for 107 out of  the  total 129 transitions. The MBPT and MCDHF/RCI2 values agree within 2\% for all the transitions. The MBPT and MCDHF/RCI2 calculations are also more accurate and reliable than the NIST values for the M1 and E2 transitions.

\section{SUMMARY}
Using the combined RCI and MBPT approach, we have performed the systematic calculations of the energies and radiative transition properties for highly charged N-like ions from \ion{Ar}{12} to \ion{Zn}{24}.  
A complete and consistent data set of energies, wavelengths, line strengths, oscillator strengths, and transition rates for all possible E1, M1, E2, and M2 transitions among the 359 levels of the $2s^2 2p^3$, $2s 2p^4$, $2p^5$, $2s^2 2p^2 3l$, $2s 2p^3 3l$, $2p^4 3l$, and $2s^2 2p^2 4l$ configurations, is provided for all the 13 ions. 
By comparing with available observations, the present MCDHF/RCI values and other elaborate/systematic theoretical calculations, high accuracy of the present MBPT results is verified, i.e., the accuracy of level energies is assessed to be within 0.1\% for most states, and is less than 0.2\% for all states; the accuracy of the line strengths is estimated to be better than  5\% for most transitions among the $n=2$ states, and for a majority of strong transitions in \ion{Fe}{20}. 
In view of the fact that our calculations are systematic and consistent, which means the unified treatment thus unified quality of data,  we anticipated that the accuracy of the line strengths is within 5\% for a majority of transitions involving the $n \geq 3$ levels, especially for the strong transitions. 
We consider that the observation uncertainty should be large for the $n \geq 3$ states included in Table~\ref{tab.lev.nistmbptlardif}, and precise experimental investigations are expected. The elaborate calculations involving higher excited states, especially for the $n = 4$ states, are also expected to further  assess the accuracy of the existing experimental and theoretical data.

Since more sufficient correlation effects have been taken into account by considering more configurations in both the \emph{N} and \emph{M} spaces, the present MBPT energies reveal an improvement in accuracy compared with the elaborate systematic calculations, such as the MCDHF/RCI work~\citep{Rynkun.2014.V100.p315} and the MBPT results~\citep{Gu.2005.V89.p267}. The present MBPT calculations agree well with the MCDHF/RCI results of \citet{Radziute.2015.V582.p61}, i.e.,  the mean energy deviation for \ion{Fe}{21} is 0.015\%. 
Meanwhile,  the present calculations have significantly increased  accurate data for  nitrogen-like ions in quantity, and may be considered as the benchmark for other calculations. The accuracy of the energy data is high enough to serve identification and interpretation of observed spectra involving the $n=3,4$ levels, for which the experimental values are largely missing.  We expect that the present data sets of high accuracy will be useful in the identification and interpretation of observed spectra, and in modeling and diagnosing of astrophysical and fusion plasmas. 

\acknowledgments
The authors express our gratitude to Dr.~MingFeng~Gu for offering guidance in using his FAC code. We acknowledge the support from the National Natural Science Foundation of China (Grant No.~21503066, No.~11504421, No.~11105015, No.~11205019, No.~11371218, No. 11274001, and No.~11474034) and the support from the Foundation for the Development of Science and Technology of Chinese Academy of Engineering Physics (Grant No.~2012B0102012). This work is also supported by NSAF under Grant No.~11076009, the Chinese Association of Atomic and Molecular Data, the Chinese National Fusion Project for ITER No. 2015GB117000, and the Swedish research council. One of the authors (KW) express his gratefully gratitude to the support from the visiting researcher program at the Fudan University.

\clearpage
\bibliographystyle{apj}
\bibliography{ref}

\onecolumn
\section*{Figure and Table}
\begin{figure*}[h]
	\epsscale{1.05}
	\plotone{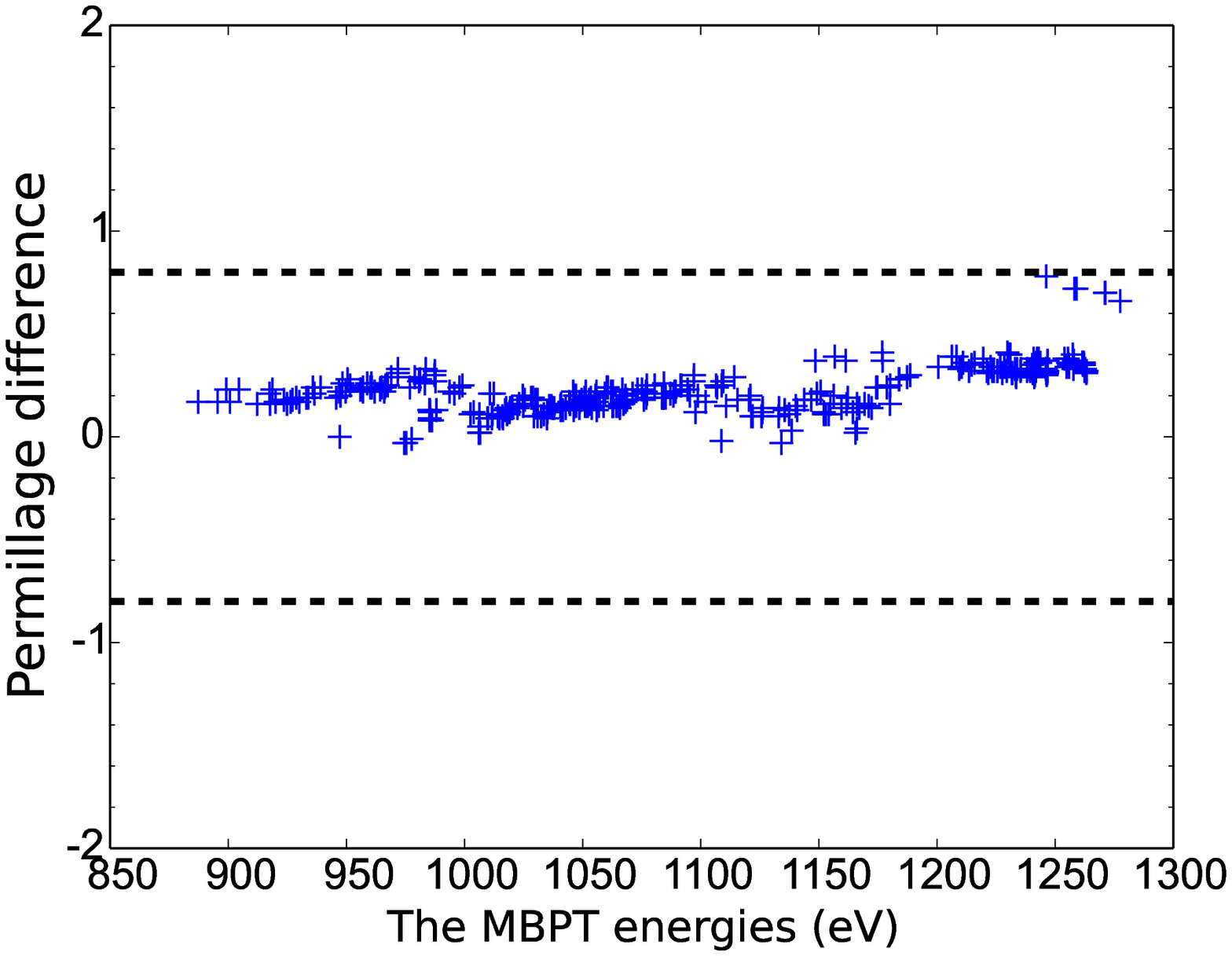}
	\caption{Permillage differences of the MCDHF/RCI values relative to the MBPT energies for the $n=3,4$ levels in \ion{Fe}{20}. Dashed lines indicate the differences of $\pm 0.08\%$. \label{fig.lev.FeXX}}

\end{figure*}

\begin{figure*}[h]
	\epsscale{1.05}
	\plotone{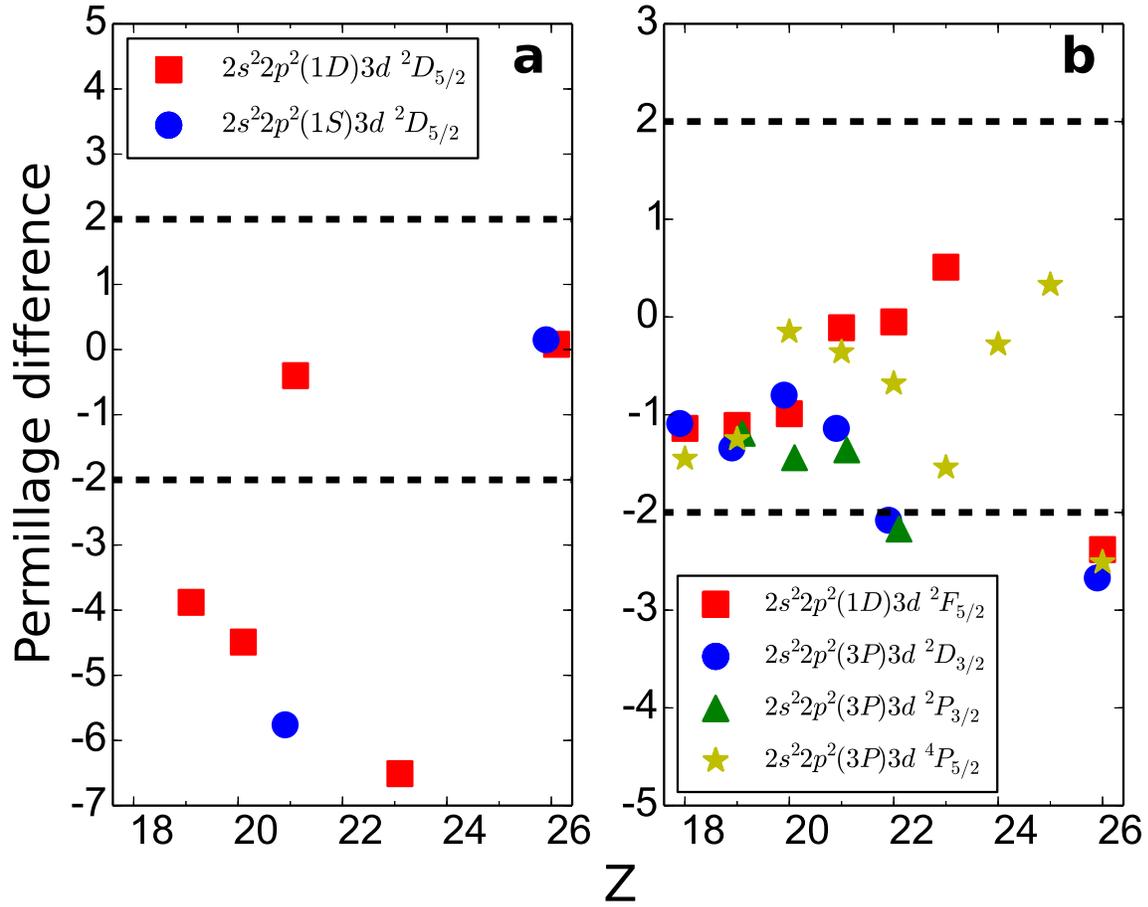}
	\caption{Permillage differences of the present MBPT energies relative to the NIST observations (a) for the $2s^2 2p^2(1D)3d\/ ~{}^2\!D_{5/2}$ and $2s^2 2p^2(1S)3d\/ ~{}^2\!D_{5/2}$ levels, and (b) for the $2s^2 2p^2(1D)3d\/~{}^2\!F_{5/2}$ and $2s^22p^2(3P)3d\/~{}^2\!D_{3/2}$ and $~{}^4\!P_{5/2}$ levels along the sequence. \label{fig.lev.nistlargedifferences}}
\end{figure*}
\begin{figure*}[t]
	\epsscale{1.05}
	\plotone{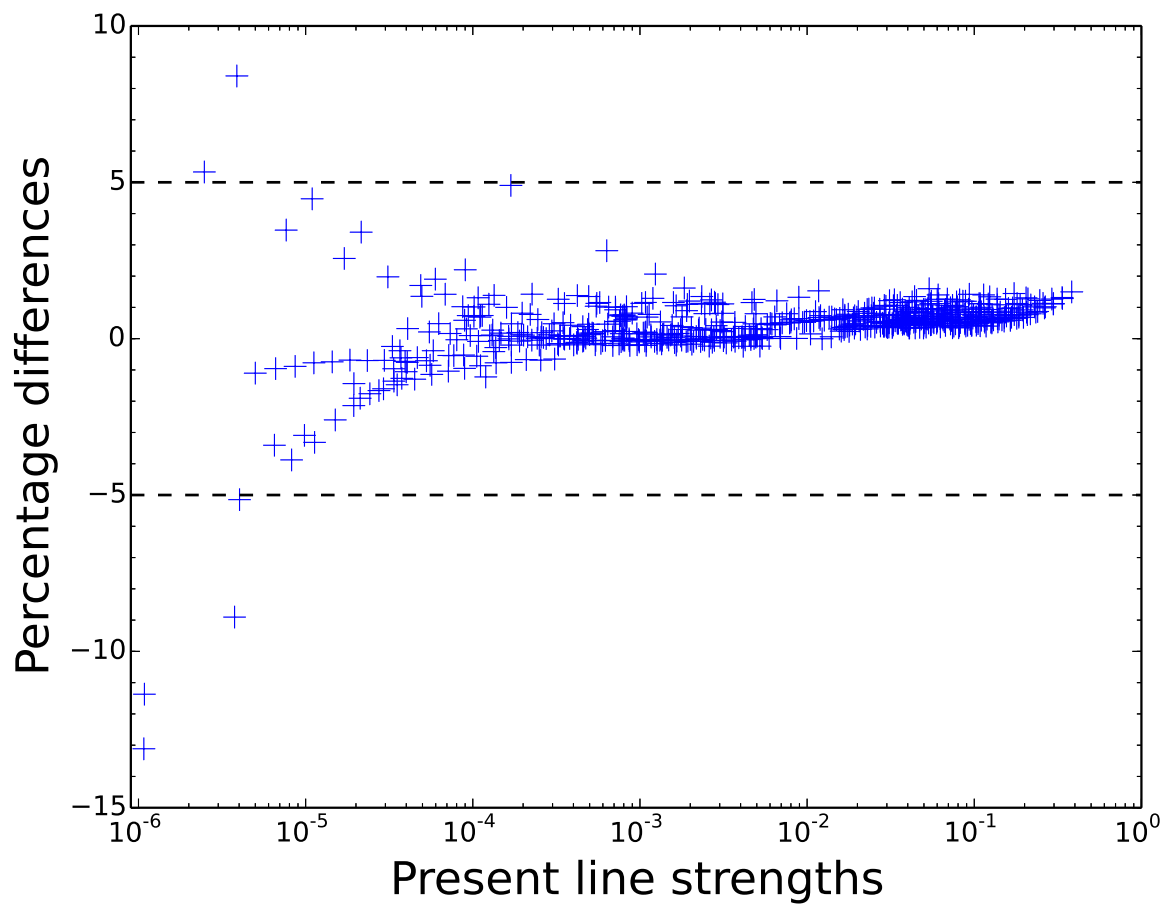}
	\caption{Percentage differences between the present MBPT line strengths and the MCDHF/RCI results  for the E1 transitions among the $n=2$ levels.  Dashed lines indicate the differences of $\pm5\%$. \label{fig.tr.e1.n2.mcdf}}
\end{figure*}

\begin{figure*}[b]
	\epsscale{1.05}
	\plotone{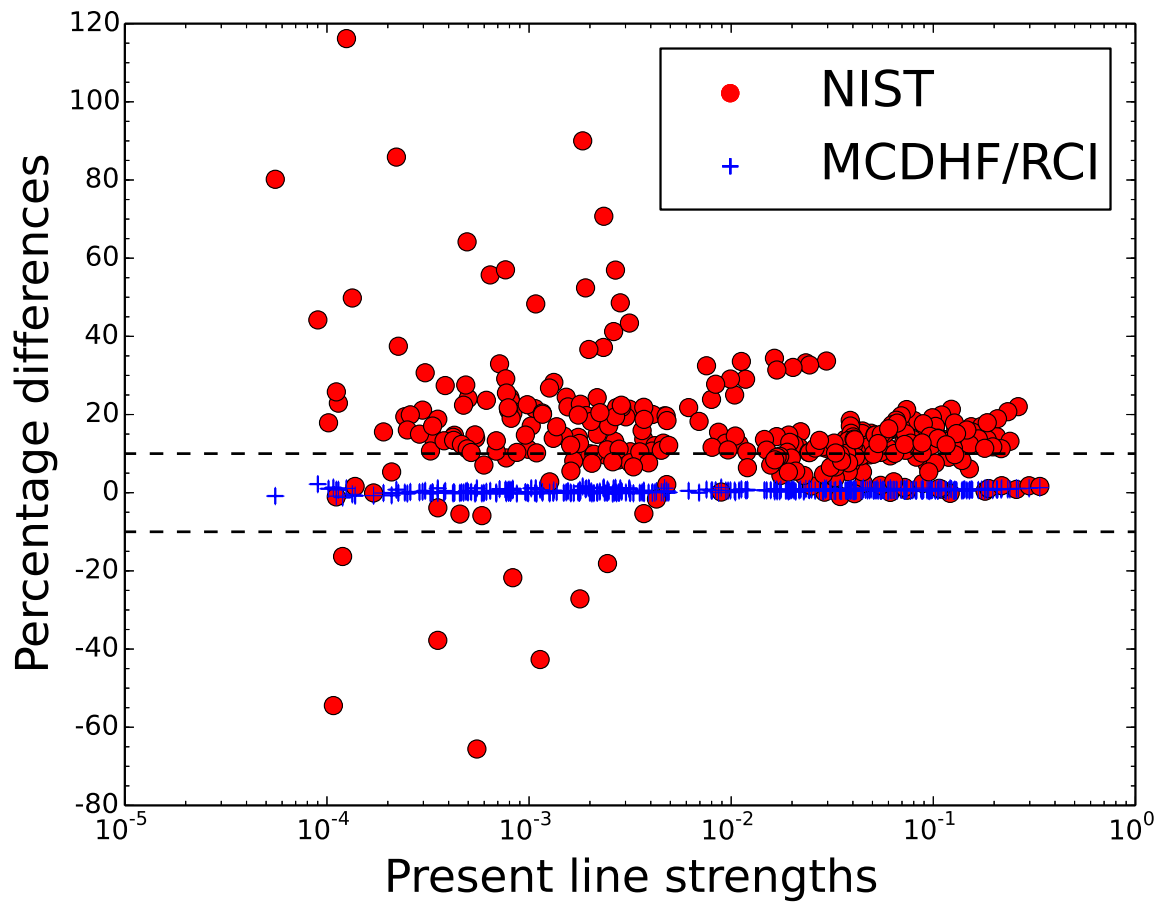}
	\caption{Percentage differences of the NIST and MCDHF/RCI2 line strengths relative to the present MBPT results for the E1 transitions among the $n=2$ levels given by the NIST ASD. Dashed lines indicate the differences of $\pm 10\%$. \label{fig.tr.e1.n2.nist}}
\end{figure*}

\begin{figure*}[b]
	\epsscale{1.05}
	\plotone{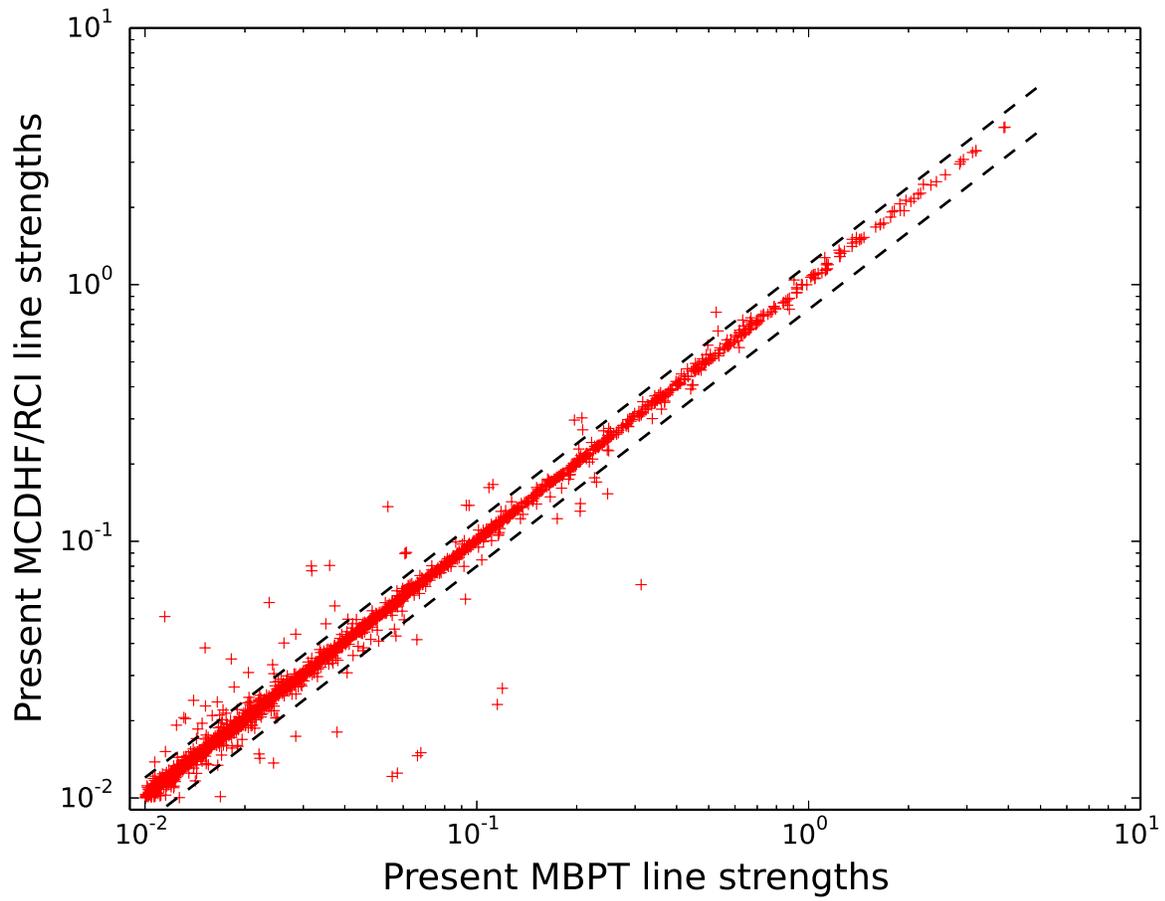}
	\caption{Comparison of the present MBPT line strengths with the MCDHF/RCI results for the strong E1 transitions involving the $n=3,4$ levels in \ion{Fe}{20}. Dashed lines indicate the differences of $\pm 20\%$. \label{fig.tr.mbpt.mcdf.FeXX}}
\end{figure*}

\clearpage
\begin{deluxetable}{llllllllr}
	\tabletypesize{\tiny}
	\tablecolumns{9}
	\tablewidth{0.0pt}
	\tablecaption{{\textnormal {Level energies (in eV) of the states in N-like ions with $Z=18-30$, as well as level designations in both the \emph{LSJ}- and \emph{jj} coupling schemes, and the dominant mixing coefficients of the $LSJ$ basis.  }\label{tab.lev.sub}}}
	\tablehead{ $Z$  &Key       &Conf       &$LSJ$     & \multicolumn{1}{c}{$jj$\tablenotemark{a,b,c}}      &$J^{\pi}$ &\multicolumn{2}{c}{Energy} & \multicolumn{1}{c}{Mixing coefficients}       \\
		\cmidrule(r){7-8}   \cmidrule(r){9-9}
		& & & & & &{NIST\tablenotemark{d}}  &{MBPT\tablenotemark{e}}  & \multicolumn{1}{c}{$LSJ$\tablenotemark{f}} \vspace{ -1.5mm} \\}
	\startdata
  $26$& $  1$& $2s^2 2p^3      $& $^4S_{3/2 }$&$2p+1(3)3                                  $&  $ 3/2^o$& $0.00000{\rm E+}00$& $0.00000{\rm E+}00$&   $ -0.94 (  1)                                           $\\
  $26$& $  2$& $2s^2 2p^3      $& $^2D_{3/2 }$&$2p-1(1)1~2p+2(4)3                         $&  $ 3/2^o$& $1.71867{\rm E+}01$& $1.71795{\rm E+}01$&   $  0.86 (  2)  -0.42 (  5)                              $\\
  $26$& $  3$& $2s^2 2p^3      $& $^2D_{5/2 }$&$2p-1(1)1~2p+2(4)5                         $&  $ 5/2^o$& $2.18373{\rm E+}01$& $2.18329{\rm E+}01$&   $  1.00 (  3)                                           $\\
  $26$& $  4$& $2s^2 2p^3      $& $^2P_{1/2 }$&$2p-1(1)1~2p+2(0)1                         $&  $ 1/2^o$& $3.22694{\rm E+}01$& $3.22817{\rm E+}01$&   $  0.99 (  4)                                           $\\
  $26$& $  5$& $2s^2 2p^3      $& $^2P_{3/2 }$&$2p+3(3)3                                  $&  $ 3/2^o$& $4.00890{\rm E+}01$& $4.01007{\rm E+}01$&   $ -0.84 (  5)  -0.48 (  2)                              $\\
  $26$& $  6$& $2s^1 2p^4      $& $^4P_{5/2 }$&$2s+1(1)1~2p+2(4)5                         $&  $ 5/2^e$& $9.33266{\rm E+}01$& $9.33152{\rm E+}01$&   $ -0.99 (  6)                                           $\\
  $26$& $  7$& $2s   2p^4      $& $^4P_{3/2 }$&$2s+1(1)1~2p-1(1)0 ~2p+3(3)3               $&  $ 3/2^e$& $1.01745{\rm E+}02$& $1.01741{\rm E+}02$&   $ -0.99 (  7)                                           $\\
  $26$& $  8$& $2s   2p^4      $& $^4P_{1/2 }$&$2s+1(1)1~2p+2(0)1                         $&  $ 1/2^e$& $1.04454{\rm E+}02$& $1.04450{\rm E+}02$&   $ -0.97 (  8)                                           $\\
\enddata
\tablenotetext{a}{The number at the end or inside of the bracket is 2$J$.}
\tablenotetext{b}{$s+ = s_{1/2}$, $p- = p_{1/2}$, $p+ = p_{3/2}$, $d- = d_{3/2}$, $d+ = d_{5/2}$, $f- = f_{5/2}$,  and $f+ = f_{7/2}$.}
\tablenotetext{c}{The number after $\pm$ is the occupation number of the corresponding sub-shell. For example, the $jj$ configuration of level 7 is $2s_{1/2} 2p_{1/2}2p_{3/2}^{3}$.}
\tablenotetext{d}{The observed energies from the NIST ASD \citep{Kramida2014}.}
\tablenotetext{e}{The present MBPT results.}
\tablenotetext{f}{The mixing coefficient of the $LSJ$ basis of the state indicated by the key in parenthesis.}\\
\tablenotetext{\space}{Note: Only the 359 levels in N-like Fe are shown here. Table~\ref{tab.lev.sub} is available online in its entirety in the home page of \emph{ApJS}. A portion is shown here for guidance regarding its form and content. }
\end{deluxetable}
\clearpage

\begin{deluxetable}{ccccccc}
	\tabletypesize{\normalsize}
	\tablecolumns{7}
	\tablewidth{0pt}
	\tablecaption{{\textnormal {Wavelengths($\lambda$, in {\AA}), line strengths(\emph{S}, in atomic units), absorption oscillator strengths (\emph{f}, dimensionless) and transition rates(\emph{A}, in s$^{-1}$) for the transitions in N-like ions with $Z=18-30$. }} \label{tab.tr.sub}}
	\tablehead{$Z$  &$j-i$   &Type  &$\lambda$    &$S$  &$f$  &$A$ \vspace{ -4mm} \\}
	\startdata
 $26$&  $ 2- 1$& M1 & $7.2170{\rm E+}02$&  $8.729{\rm E-}01$&  $1.223{\rm E-}06$&  $1.566{\rm E+}04$\\
 $26$&  $ 2- 1$& E2 & $7.2170{\rm E+}02$&  $5.198{\rm E-}04$&  $5.804{\rm E-}11$&  $7.433{\rm E-}01$\\
 $26$&  $ 3- 1$& M1 & $5.6788{\rm E+}02$&  $5.106{\rm E-}02$&  $9.091{\rm E-}08$&  $1.254{\rm E+}03$\\
 $26$&  $ 3- 1$& E2 & $5.6788{\rm E+}02$&  $1.066{\rm E-}03$&  $2.443{\rm E-}10$&  $3.369{\rm E+}00$\\
 $26$&  $ 3- 2$& M1 & $2.6644{\rm E+}03$&  $1.780{\rm E+}00$&  $6.752{\rm E-}07$&  $4.230{\rm E+}02$\\
 $26$&  $ 3- 2$& E2 & $2.6644{\rm E+}03$&  $1.909{\rm E-}03$&  $4.237{\rm E-}12$&  $2.654{\rm E-}03$\\
 $26$&  $ 4- 1$& M1 & $3.8407{\rm E+}02$&  $1.316{\rm E-}01$&  $3.464{\rm E-}07$&  $3.133{\rm E+}04$\\
 $26$&  $ 4- 1$& E2 & $3.8407{\rm E+}02$&  $1.016{\rm E-}04$&  $7.528{\rm E-}11$&  $6.808{\rm E+}00$\\
\enddata
		\tablenotetext{\space}{Note: Only transitions among the $n=2$ levels in \ion{Fe}{20} are shown here. Table~\ref{tab.tr.sub} is available online in its entirety in the home page of \emph{ApJS}. A portion is shown here for guidance regarding its form and content.}
\end{deluxetable}

\begin{deluxetable}{lccccc}
	
	\tablecolumns{7}
	\tablewidth{0pt}
	\tablecaption{{\textnormal { Level energies (in eV) for the $n=3,4$ levels in Fe XX. }}\label{tab.lev.FeXX}}
	\tablehead{ \multirow{2}{0.2in}{\large Level \vspace{ -7mm} }   & \multicolumn{5}{c}{\large Energy (eV) \vspace{ 1mm}  }\\
		\cmidrule(r){2-6} \vspace{ -5mm} 
		&NIST$^a$ &CHIANTI$^b$ &MBPT$^c$ &MCDHF/RCI$^d$ &AS$^e$\\}
	\startdata
	$2s^{2}2p^{2}(^3P)3s       \   {}^{4}P_{1/2 }$    &  \nodata&  887.355& 887.256 & 887.403 & 889.851 \\
	$2s^{2}2p^{2}(^3P)3s       \   {}^{4}P_{3/2 }$    &  \nodata&  895.581& 895.508 & 895.656 & 897.924 \\
	$2s^{2}2p^{2}(^3P)3s       \   {}^{2}P_{1/2 }$    &  \nodata&  \nodata& 899.163 & 899.367 & 901.900 \\
	$2s^{2}2p^{2}(^3P)3s       \   {}^{4}P_{5/2 }$    &  \nodata&  899.702& 900.731 & 900.888 & 903.450 \\
	$2s^{2}2p^{2}(^3P)3s       \   {}^{2}P_{3/2 }$    &  \nodata&  \nodata& 904.521 & 904.731 & 907.556 \\
	$2s^{2}2p^{2}(^3P)3p       \   {}^{4}D_{1/2 }$    &  \nodata&  \nodata& 912.215 & 912.364 & 914.791 \\
	$2s^{2}2p^{2}(^1D)3s       \   {}^{2}D_{5/2 }$    &  \nodata&  \nodata& 917.442 & 917.635 & 920.142 \\
	$2s^{2}2p^{2}(^3P)3p       \   {}^{4}D_{3/2 }$    &  \nodata&  916.602& 917.687 & 917.834 & 920.134 \\
\enddata
	\tablenotetext{a}{The observed energies from the NIST ASD~\citep{Kramida2014}. }
	\tablenotetext{b}{The observed energies from the Chianti database~\citep{Landi.2013.V763.p86,Dere.1997.V125.p149}.}
	\tablenotetext{c}{The present MBPT energies.}
	\tablenotetext{d}{The present MCDHF/RCI energies.}
	\tablenotetext{e}{The energies calculated by~\citet{Witthoeft.2007.V466.p763} using the AS code.}
\end{deluxetable}

\begin{deluxetable}{lllllc}
	\tabletypesize{\large}
	\tablecolumns{9}
	\tablewidth{0pt}
	\tablecaption{{\textnormal {Comparisons of the levels with differences larger than 0.2\% between our MBPT energies and the experimental values of the NIST ASD. }} \label{tab.lev.nistmbptlardif}}
	\tablehead{\colhead{\emph{Z}} &\colhead{Key\tablenotemark{a}}  &\colhead{State}  & \multicolumn{2}{c}{ Energy (eV)} &\colhead{Difference (\%)} \\
		\cmidrule(r){4-5}
		& & &\colhead{NIST\tablenotemark{b}} &\colhead{MBPT\tablenotemark{c}}  \vspace{ -6mm} \\}
	\startdata
	18 & 33 &  $2s^2 2p^2 (^3P)3p \ {}^{4}S_{3/2}$ &    379.27  &  376.264&   0.79\\
	18 & 41 &  $2s^2 2p^2 (^1D)3p \ {}^{2}D_{5/2}$ &    387.43  &  385.478&   0.50\\
	18 & 42 &  $2s^2 2p^2 (^1D)3p \ {}^{2}P_{1/2}$ &    388.54  &  387.446&   0.28\\
	18 & 43 &  $2s^2 2p^2 (^1D)3p \ {}^{2}P_{3/2}$ &    389.62  &  388.648&   0.25\\
	18 & 63 &  $2s^2 2p^2 (^3P)3d \ {}^{2}D_{5/2}$ &    400.07  &  399.251&   0.20\\
	18 & 67 &  $2s^2 2p^2 (^1D)3d \ {}^{2}G_{7/2}$ &    404.004 &  401.334&   0.66\\
	19 & 63 &  $2s^2 2p^2 (^3P)3d \ {}^{2}D_{5/2}$ &    458.5   &  457.275&   0.27\\
	19 & 70 &  $2s^2 2p^2 (^1D)3d \ {}^{2}D_{5/2}$ &    463.2   &  461.403&   0.39\\
\enddata
		\tablenotetext{a}{The index number of the level given in Table~\ref{tab.lev.sub}.}
		\tablenotetext{b}{The NIST recommended energies \citep{Kramida2014}.}
		\tablenotetext{c}{The present MBPT energies.}
\end{deluxetable}
\clearpage
\begin{deluxetable}{ccccccc}
	
	\tablecolumns{7}
	\tablewidth{0pt}
	\tablecaption{{\textnormal {Comparisons of the line strengths with differences large than 10\% between our MBPT results and the NIST values for transitions among the $n=2$ levels. The MCDHF/RCI2 results are also listed for comparison.}}\label{tab.tr.e1.n2.nist.largediff}}
	\tablehead{ \scriptsize $Z$   &\scriptsize Transition & \multicolumn{3}{c}{\scriptsize Line Strengths} &\multicolumn{2}{c}{\scriptsize{Difference (\%)}}\\
		\cmidrule(r){3-5}   \cmidrule(r){6-7}
		&&\scriptsize MBPT &\scriptsize MCDHF/RCI2 &\scriptsize NIST &\scriptsize{MBPT\&MCDHF/RCI2} &\scriptsize{MBPT\&NIST}\vspace{ -4mm} \\}
	\startdata
$21$&       $2s    2p^{4} \ {}^{4}\!P_{5/2} - 2s^{2}2p^{3} \ {}^{4}\!S_{3/2}$&     $1.429{\rm E-}01$&	$1.441{\rm E-}01$&	$1.600{\rm E-}01$&	$0.1 $&$12 $\\
$21$&       $2s    2p^{4} \ {}^{4}\!P_{5/2} - 2s^{2}2p^{3} \ {}^{2}\!D_{3/2}$&     $1.527{\rm E-}03$&	$1.529{\rm E-}03$&	$1.900{\rm E-}03$&	$0.1 $&$24 $\\
$21$&       $2s    2p^{4} \ {}^{4}\!P_{5/2} - 2s^{2}2p^{3} \ {}^{2}\!P_{3/2}$&     $7.145{\rm E-}04$&	$7.211{\rm E-}04$&	$9.500{\rm E-}04$&	$0.9 $&$33 $\\
$21$&       $2s    2p^{4} \ {}^{4}\!P_{3/2} - 2s^{2}2p^{3} \ {}^{4}\!S_{3/2}$&     $9.674{\rm E-}02$&	$9.757{\rm E-}02$&	$1.120{\rm E-}01$&	$0.9 $&$16 $\\
$21$&       $2s    2p^{4} \ {}^{4}\!P_{3/2} - 2s^{2}2p^{3} \ {}^{2}\!P_{3/2}$&     $1.326{\rm E-}03$&	$1.333{\rm E-}03$&	$1.700{\rm E-}03$&	$0.5 $&$28 $\\
$21$&       $2s    2p^{4} \ {}^{4}\!P_{1/2} - 2s^{2}2p^{3} \ {}^{4}\!S_{3/2}$&     $4.921{\rm E-}02$&	$4.963{\rm E-}02$&	$5.670{\rm E-}02$&	$0.9 $&$15 $\\
$21$&       $2s    2p^{4} \ {}^{4}\!P_{1/2} - 2s^{2}2p^{3} \ {}^{2}\!P_{1/2}$&     $3.323{\rm E-}04$&	$3.326{\rm E-}04$&	$3.800{\rm E-}04$&	$0.1 $&$14 $\\
$21$&       $2s    2p^{4} \ {}^{2}\!D_{3/2} - 2s^{2}2p^{3} \ {}^{4}\!S_{3/2}$&     $2.972{\rm E-}04$&	$2.979{\rm E-}04$&	$3.600{\rm E-}04$&	$0.3 $&$21 $\\
\enddata
\end{deluxetable}
\clearpage

\begin{deluxetable}{ccccccc}
	
	\tablecolumns{7}
	\tablewidth{0pt}
	\tablecaption{{\textnormal {Comparisons of the MCDHF/RCI line strengths and the NIST values relative to the present MBPT  results for transitions involving the $n=3,4$ levels.}}\label{tab.tr.FeXX.n3.nist}}
	\tablehead{ \scriptsize $Z$   &\scriptsize Transition & \multicolumn{3}{c}{\scriptsize Line Strengths } &\multicolumn{2}{c}{\scriptsize{Difference (\%)}}\\
		\cmidrule(r){3-5}   \cmidrule(r){6-7}
		&&\scriptsize MBPT &\scriptsize MCDHF/RCI &\scriptsize NIST &\scriptsize{MBPT\&MCDHF/RCI} &\scriptsize{MBPT\&NIST}\vspace{ -4mm} \\}
	\startdata
$26$&	  $2s^2 2p^ 2(^3P)3d \ {}^{4}\!P_{3/2} - 2s^2 2p^3 \ {}^{4}S_{3/2}$&   $8.230{\rm E-}02$&      $8.25{\rm E-}02$&       $8.70{\rm E-}02$&  $  0.2$&  $6  $\\      
$26$&	  $2s^2 2p^ 2(^3P)3d \ {}^{4}\!P_{3/2} - 2s^2 2p^3 \ {}^{2}D_{5/2}$&   $3.902{\rm E-}03$&      $3.92{\rm E-}03$&       $3.60{\rm E-}03$&  $  0.6$&  $-8 $\\      
$26$&	  $2s^2 2p^ 2(^3P)3d \ {}^{2}\!D_{3/2} - 2s^2 2p^3 \ {}^{4}S_{3/2}$&   $9.373{\rm E-}04$&      $9.62{\rm E-}04$&       $8.50{\rm E-}04$&  $  2.6$&  $-9 $\\      
$26$&	  $2s^2 2p^ 2(^3P)3d \ {}^{2}\!D_{3/2} - 2s^2 2p^3 \ {}^{2}D_{5/2}$&   $4.479{\rm E-}03$&      $4.51{\rm E-}03$&       $4.80{\rm E-}03$&  $  0.7$&  $7  $\\      
$26$&	  $2s^2 2p^ 2(^3P)3d \ {}^{2}\!D_{3/2} - 2s^2 2p^3 \ {}^{2}P_{1/2}$&   $4.412{\rm E-}02$&      $4.43{\rm E-}02$&       $4.00{\rm E-}02$&  $  0.4$&  $-9 $\\      
$26$&	  $2s^2 2p^ 2(^3P)3d \ {}^{2}\!D_{3/2} - 2s^2 2p^3 \ {}^{2}P_{3/2}$&   $5.593{\rm E-}03$&      $5.58{\rm E-}03$&       $5.50{\rm E-}03$&  $ -0.2$&  $-2 $\\      
$26$&	  $2s^2 2p^ 2(^3P)3d \ {}^{2}\!D_{5/2} - 2s^2 2p^3 \ {}^{4}S_{3/2}$&   $7.721{\rm E-}03$&      $7.89{\rm E-}03$&       $6.80{\rm E-}03$&  $  2.2$&  $-12$\\      
$26$&	  $2s^2 2p^ 2(^3P)3d \ {}^{2}\!D_{5/2} - 2s^2 2p^3 \ {}^{2}D_{3/2}$&   $1.082{\rm E-}01$&      $1.09{\rm E-}01$&       $1.00{\rm E-}01$&  $  0.6$&  $-8 $\\
\enddata
\end{deluxetable}

\begin{deluxetable}{cccccccc}
	
	\tablecolumns{8}
	\tablewidth{0pt}
	\tablecaption{{\textnormal {Comparisons of the MCDHF/RCI2 line strengths 
 and the NIST values relative to the present MBPT  results 
  for the M1 and E2 transitions among the $n=2$ levels.}} \label{tab.tr.m1.e2.n2.nist.largediff}}
	\tablehead{ \scriptsize $Z$   &\scriptsize Transition &\scriptsize Type & \multicolumn{3}{c}{\scriptsize Line strengths } &\multicolumn{2}{c}{\scriptsize{Difference (\%)}}\\
		\cmidrule(r){4-6}   \cmidrule(r){7-8}
		&&&\scriptsize MBPT &\scriptsize MCDHF/RCI2 &\scriptsize NIST &\scriptsize{MBPT\&MCDHF/RCI2} &\scriptsize{MBPT\&NIST}\vspace{ -4mm} \\}
	\startdata
$19$&  $2s^{2}2p^{3} \ {}^{2}\!D_{3/2} -    2s^{2}2p^{3} \ {}^{4}\!S_{3/2}$&     M1 &  $3.312{\rm E-}02$& $3.303{\rm E-}02$&	$3.190{\rm E-}02$&    $ -0.3$&    $-4 $\\
$19$&  $2s^{2}2p^{3} \ {}^{2}\!P_{1/2} -    2s^{2}2p^{3} \ {}^{4}\!S_{3/2}$&     M1 &  $1.642{\rm E-}02$& $1.641{\rm E-}02$&	$1.600{\rm E-}02$&    $ -0.1$&    $-3 $\\
$19$&  $2s^{2}2p^{3} \ {}^{2}\!P_{3/2} -    2s^{2}2p^{3} \ {}^{4}\!S_{3/2}$&     M1 &  $5.796{\rm E-}02$& $5.816{\rm E-}02$&	$5.680{\rm E-}02$&    $ 0.3 $&    $-2 $\\
$21$&  $2s^{2}2p^{3} \ {}^{2}\!D_{3/2} -    2s^{2}2p^{3} \ {}^{4}\!S_{3/2}$&     M1 &  $1.048{\rm E-}01$& $1.045{\rm E-}01$&	$1.186{\rm E-}01$&    $ -0.3$&    $13 $\\
$21$&  $2s^{2}2p^{3} \ {}^{2}\!D_{5/2} -    2s^{2}2p^{3} \ {}^{2}\!D_{3/2}$&     M1 &  $2.055{\rm E+}00$& $2.064{\rm E+}00$&	$2.020{\rm E+}00$&    $ 0.5 $&    $-2 $\\
$21$&  $2s^{2}2p^{3} \ {}^{2}\!D_{5/2} -    2s^{2}2p^{3} \ {}^{2}\!D_{3/2}$&     E2 &  $3.768{\rm E-}03$&           \nodata&	$4.500{\rm E-}03$&    \nodata&    $19 $\\
$21$&  $2s^{2}2p^{3} \ {}^{2}\!P_{1/2} -    2s^{2}2p^{3} \ {}^{4}\!S_{3/2}$&     M1 &  $3.253{\rm E-}02$& $3.250{\rm E-}02$&	$3.522{\rm E-}02$&    $ -0.1$&    $8  $\\
$21$&  $2s^{2}2p^{3} \ {}^{2}\!P_{1/2} -    2s^{2}2p^{3} \ {}^{2}\!D_{3/2}$&     M1 &  $1.722{\rm E-}01$& $1.731{\rm E-}01$&	$1.857{\rm E-}01$&    $ 0.5 $&    $8  $\\
\enddata
\end{deluxetable}
\end{document}